\documentclass[twocolumn,
showpacs,
aps,pra,
superscriptaddress
]{revtex4-1}

\usepackage{ucs}
\usepackage{graphicx}
\usepackage{amssymb}
\usepackage{amsmath}
\usepackage[usenames,dvipsnames]{color}
\usepackage{bm}
\usepackage{epstopdf}

\newcommand{\slfrac}[2]{\left.#1\middle/#2\right.}

\DeclareMathOperator*{\To}{\to}

\begin{document}
\pacs{67.85.Lm, 67.85.De, 03.75.Ss}
\title{Soliton core filling in superfluid Fermi gases with spin-imbalance.}
\author{G. Lombardi}
\email{giovanni.lombardi@uantwerpen.be}
\affiliation{TQC, Universiteit Antwerpen, Universiteitsplein 1, B-2610 Antwerpen, Belgium}
\author{W. Van Alphen}
\affiliation{TQC, Universiteit Antwerpen, Universiteitsplein 1, B-2610 Antwerpen, Belgium}
\author{S. N. Klimin}
\affiliation{TQC, Universiteit Antwerpen, Universiteitsplein 1, B-2610 Antwerpen, Belgium}
\affiliation{Department of Theoretical Physics, State University of Moldova, 2009 Chi\unichar{537}in\unichar{259}u, Moldova}
\author{J. Tempere}
\affiliation{TQC, Universiteit Antwerpen, Universiteitsplein 1, B-2610 Antwerpen, Belgium}
\affiliation{Lyman Laboratory of Physics, Harvard University, Cambridge, MA 02138, USA}

\begin{abstract}
In this paper the properties of dark solitons in superfluid Fermi gases with spin-imbalance are studied by means of a recently developed effective field theory [S. N. Klimin \textit{et al.}, Eur. Phys. J. B {\bf{88}}, 122 (2015)] suitable to describe the BEC-BCS crossover in ultracold gases in an extended range of temperatures as compared to the usual Ginzburg-Landau treatments.
The spatial profiles for the total density and for the density of the excess-spin component, and the changes of their properties across the BEC-BCS crossover are examined in different conditions of temperature and imbalance. The presence of population imbalance is shown to strongly affect the structure of the soliton excitation by filling its core with unpaired atoms. This in turn influences the dynamical properties of the soliton since the additional particles in the core have to be dragged along thus altering the effective mass.
\end{abstract}

\maketitle

\section{Introduction}
Solitons are nonlinear localized excitations that propagate trough a medium without changing their shape: these properties make them extremely valuable in the study of the interplay between coherence and interactions and hence they were examined in an extremely broad set of physical systems, ranging from macroscopic solitary waves in water canals, to pulses in optical fibers \citep{EXPKrokelHalas}, to excitations in conducting polymers \citep{THHeeger}.\\
In the field of ultracold atoms a great deal of attention has been focused on dark solitons: these appear as localized density dips propagating on a stable background and are characterized by a phase jump across the density minimum. While the first experimental studies of dark solitons in BEC condensates date back to the first decade of the 2000s \citep{EXPDenschlag,EXPBurgerBongs,EXPAndersonHaljian,EXPBeckerStellmer} their investigation in fermionic ultracold gases is more recent. In 2013 an experiment at MIT \citep{EXPYefsah} detected long-lived solitary waves that were initially identified as dark solitons; a later paper by the same research group \citep{EXPKu} explained the surprising long life and high effective mass of the observed excitations by correctly reinterpreting them as solitonic vortices: products of the decay of dark solitons via the snake instability \citep{THCetoliBrand,THMunozBrand}, already predicted and experimentally detected \citep{EXPDonadello} in BEC condensates. 
The main technique employed to create solitons in ultracold gases is phase imprinting \citep{EXPBurgerBongs,EXPDenschlag,EXPYefsah,EXPKu,EXPAndersonHaljian, EXPBeckerStellmer}. The starting point is an ultracold gas in which the interparticle interaction can be tuned by using a Feshbach resonance. A laser beam is shone on one half of the atomic cloud, thus locally changing the phase of the order parameter. The soliton is then observed via absorption imaging after a time-of-flight expansion \citep{EXPBurgerBongs,EXPDenschlag,EXPAndersonHaljian, EXPBeckerStellmer}. In the case of Fermi superfluids, during the time-of-flight expansion a rapid ramp towards the BEC side of the resonance is also employed \citep{EXPYefsah,EXPKu}. In 2014 \citep{EXPSachaDelande} a technique was proposed in order to create stable dark solitons in fermionic systems by applying the phase imprinting method to just one of the two spin-components of the gas.
\\From the theoretical point of view, solitons in BEC-BCS crossover superfluids have been widely studied by solving the Bogoliubov-de Gennes equations \citep{THSpuntarelli,THEfimkin,THScottDalfovo,THAntezzaDalfovo} or -- limited to the deep BEC side of the crossover -- the Gross-Pitaevskii equation \citep{THKonotopPitaevskii, THFrantzeskakis}. In this article we are going to perform the study of solitons in imbalanced fermionic superfluids in the context of the effective field theory (EFT) presented in \citep{THKTLDEpjB} and already employed for the examination of the population-balanced case \citep{THKTDPrA}. This theory has the advantage of being applicable to the whole range of temperatures below the critical temperature $T_c$ (correctly retrieving the results of the Ginzburg-Landau treatment in the vicinity of $T_c$ \citep{THRanderiaSaDeMelo} and of the effective theories developed specifically for $T=0$ \citep{THMariniPistolesiStrinati,THDienerSensarmaRanderia, THHuang}) and of being much less demanding than the BdG approach from a computational point of view.
In this paper we study dark solitons in superfluid Fermi gases with the addition of spin-imbalance, i.e. the possibility of having unequal populations of spin-up and spin-down particles in the ultracold gas. Imbalance in ultracold Fermi systems was first experimentally engineered in 2006 \citep{EXPZwierlein,EXPPartridge} but, because of its effects on the pairing mechanism, it has been a focus of major theoretical attention starting from the seminal works on critical fields in superconductors \citep{THClogston,THChandrasekhar}, up to more recent papers on ultracold fermions \citep{THTKDPrA79}.
Dark solitons, like vortices, are characterized by a dip in the order parameter's profile: we will show that, in presence of population imbalance, this dip provides a favorable place where to accommodate the excess component particles. A travelling soliton has to transport these excess particles: we compute how this affects the basic dynamical quantities of the system.  
\\The present article is organized as follows: in Section \ref{model} we give a brief sketch of the theoretical model employed to examine our system \citep{THKTDPrA,THKTLDEpjB}. Section \ref{dssolution} is dedicated to the solution of the equations of motion and the definition of some of the relevant quantities that will be analyzed in the following; in Section \ref{spatialprofiles} we study the various aspects related to the filling of the soliton in presence of imbalance and consider how the range of applicability of the EFT is affected by temperature and imbalance; finally the conclusions of our investigation are given in Section \ref{conclusions}.

\section{Model}\label{model}

The formalism employed in this treatment is the one presented in \citep{THKTDPrA}. Assuming slow variations of the superfluid order parameter $\Psi$ in both time and space, the effective action for the system in the natural units $\hbar=1$, $2m=1$, $E_F=1$, $v_F=\hbar k_F/m=2$, is
\begin{equation}
S(\beta)=\int_0^\beta\mathrm{d}\tau\int\mathrm{d}\bm{r}\left[\frac{D}{2}\left(\bar{\Psi}\frac{\partial \Psi}{\partial\tau}-\frac{\partial\bar{\Psi}}{\partial \tau}\Psi\right)+\mathcal{H}\right] \,,\label{Sbeta}
\end{equation}
where $\beta$ is the inverse temperature and the Hamiltonian $\mathcal{H}$ is given by
\begin{equation}
\mathcal{H}=\Omega_s+\frac{C}{2m}\left|\nabla_{\bm{r}}\Psi\right|^2 -\frac{E}{2m}\left(\nabla_{\bm{r}}\left|\Psi\right|\right)^2 \,. \label{H}
\end{equation}
The first term appearing in this formula is the thermodynamic potential $\Omega_s$ that reads
\begin{align}
\Omega_s=&-\int\frac{\mathrm{d}\bm{k}}{\left(2\pi\right)^3}\Bigg[\frac{1}{\beta}\log\left[2\cosh\left(\beta E_{\bm{k}}\right)+2\cosh\left(\beta \zeta\right)\right]+\nonumber\\&-\xi_{\bm{k}}-\frac{m\left|\Psi\right|^2}{k^2}\Bigg]-\frac{m\left|\Psi\right|^2}{4\pi a_s} \,.
\end{align}
Here $\xi_{\bm{k}}=\frac{k^2}{2m}-\mu$ is the dispersion relation for a free fermion, $E_{\bm{k}}=\sqrt{\xi_{\bm{k}}+|\Psi|^2}$ is the Bogoliubov excitation energy and $a_s$ is the $s-$wave scattering length that determines the strength and sign of the contact interaction. Since we want to consider a system with spin-imbalance we introduce different chemical potentials $\mu_\uparrow$ and $\mu_\downarrow$ for fermions with spin up and down respectively. These are in turn linked to the quantities $\mu$ and $\zeta$ by the relations
$\mu=(\mu_\uparrow+\mu_\downarrow)/2$ and $\zeta=(\mu_\uparrow-\mu_\downarrow)/2$.\\
The other coefficients appearing in \eqref{Sbeta} and \eqref{H} are defined as
\begin{align}
C=&\int\frac{\mathrm{d}\bm{k}}{(2\pi)^3}\frac{k^2}{3m}f_2(\beta, E_{\bm{k}},\zeta)\,,\label{C}\\
D=&\int\frac{\mathrm{d}\bm{k}}{(2\pi)^3}\frac{\xi_{\bm{k}}}{\left|\Psi\right|^2}\left[f_1(\beta, \xi_{\bm{k}},\zeta)-f_1(\beta, E_{\bm{k}},\zeta)\right]\,, \label{D} \\
E=&\int\frac{\mathrm{d}\bm{k}}{(2\pi)^3}\frac{k^2}{3m}2\xi_{\bm{k}}^2f_4(\beta, E_{\bm{k}},\zeta)\, , \label{E}
\end{align}
where the functions $f_s(\beta,x,\zeta)$ introduced in the last set of expressions are the solutions of
\begin{equation}
f_s(\beta,x,\zeta)\equiv\frac{1}{\beta}\sum_{n}\frac{1}{\left[\left(\omega_n-\mathrm{i}\zeta\right)^2 +x^2\right]^s}
\end{equation}
with the fermionic Matsubara frequencies $\omega_n=(2n+1)\pi / \beta$. The analytic expression for $f_1(\beta,x,\zeta)$ and the recursion relation that links it to the other $f_s(\beta,x,\zeta)$ are given in \citep{THKTLDEpjB}.\\
As already mentioned, the two main characteristics of a soliton are a sharp phase gradient and a dip in the spatial profile of the order parameter, therefore we need a formalism that can properly describe both aspects. This can be obtained by separating these two contributions as
\begin{equation}
\Psi(\bm{r},t)=\left|\Psi(\bm{r},t)\right|e^{\mathrm{i}\theta(\bm{r},t)}.
\end{equation}
Moreover we can highlight the change in the amplitude by defining the modulus of the order parameter as its bulk value (that can be obtained from the solution of the gap equation) multiplied with the amplitude modulation function $a$ as
$\left|\Psi(\bm{r},t)\right|=\left|\Psi_\infty\right|a(\bm{r},t)$.
From this it becomes immediately clear that, for a localized excitation such as a soliton, $a(r\rightarrow\infty ,t)\equiv a_\infty=1$.
The real-time Lagrangian for the system is given by
\begin{align}
\mathcal{L}=-\kappa(a)a^2 \frac{\partial \theta}{\partial t}-\mathcal{H} \, ,\label{lagrangian}
\end{align}
with the real-time Hamiltonian
\begin{align}
\mathcal{H}=\Omega_s(a)-\Omega_s(a_\infty)+\frac{\rho_{qp}(a)}{2}\left(\nabla_{\bm{r}}a\right)^2+\frac{\rho_{sf}(a)}{2}\left(\nabla_{\bm{r}}\theta\right)^2\, .
\end{align}
The coefficient $\kappa$ of the term with the time derivative, the quantum pressure $\rho_{qp}$ and the superfluid density $\rho_{sf}$ are related to the coefficients $C$, $D$ and $E$ (eqs. \eqref{C}, \eqref{D}, \eqref{E}) by
\begin{align}
\kappa (a)=&D\left|\Psi_\infty\right|^2 \,,\label{kappa}\\
\rho_{sf}(a)=&\frac{C}{m}\left|\Psi\right|^2 \,,\label{rhosf}\\
\rho_{qp}(a)=&\frac{C-4\left|\Psi\right|^2E}{m}\left|\Psi_\infty\right|^2 \, .\label{rhoqp}
\end{align}
\section{Dark soliton solution} \label{dssolution}
Imposing the condition
\begin{equation}
a(x,t)\longrightarrow a(x-v_S t)\,,\qquad \theta(x, t)\longrightarrow \theta(x- v_S t) \label{solitondependence}
\end{equation} requiring that the soliton propagates with constant velocity $v_S$ along the direction $\hat{x}$, and demanding that the total change of phase across the soliton
\begin{equation}
\Delta \theta = \theta(x\rightarrow\infty)-\theta(x\rightarrow-\infty)
\end{equation}
remains finite, we can solve the Lagrange equations for \eqref{lagrangian} for $a(x)$ and $\theta(x)$ as in \citep{THKTDPrA}.
The spatial dependence of the phase is described by
\begin{equation}
\theta(x)=v_S\int_{-\infty}^{x}\mathrm{d}x'\,\frac{\kappa(a(x'))a(x')^2-\kappa(a_\infty)}{\rho_{sf}(a(x'))} \,.
\end{equation}
From the Lagrange equation for $a$ we instead obtain the following relation that gives the position (i.e. the distance from the soliton center) for each value of the amplitude
\begin{equation}
x=\pm\frac{1}{\sqrt{2}}\int_{a_0}^a \mathrm{d}a' \sqrt{\frac{\rho_{qp}(a')}{X(a')-v_S^2Y(a')}}\,.
\end{equation}
In this expression the functions $X(a)$ and $Y(a)$ are defined as
\begin{align}
X(a)&\equiv \Omega_s(a)-\Omega_s(a_\infty)\,,\\
Y(a)&\equiv \frac{\left[\kappa(a)a^2 -\kappa_\infty\right]^2}{2\rho_{sf}(a)}\, ,
\end{align}
and the amplitude at the center of the soliton $a_0\equiv a(x=0)$ is easily determined as the solution of $X(a_0)-v_S^2Y(a_0)=0$.\\
At a later stage of the present paper we are going to briefly consider a few aspects related to the dynamics of the soliton, therefore we give here the definitions for the soliton's momentum and energy. 
\begin{align}
\mathcal{P}_S(v_S)&=2 \sqrt{2} \,v_S \int_{a_0}^1 \mathrm{d}a \frac{\sqrt{\rho_{qp}(a)}Y(a)}{\sqrt{X(a)-v_S^2Y(a)}}-\pi\kappa_\infty \,,\label{Ps}\\
\mathcal{E}_S(v_S)&=2 \sqrt{2} \int_{a_0}^1 \mathrm{d}a \frac{\sqrt{\rho_{qp}(a)}X(a)}{\sqrt{X(a)-v_S^2Y(a)}} \,.\label{Es}
\end{align}
For a detailed derivation of the above expressions we address again the reader to \citep{THKTDPrA}.\\
Before presenting our results it is necessary to make a further consideration. Since one fundamental element in the derivation of the EFT that we are employing is a gradient expansion of the pairing field up to second order in spatial and time gradients, we have to carefully examine where we must consider the coordinate dependence of the coefficients $C$, $D$ and $E$ and where doing so would lead us beyond the limits of our approximation. From \eqref{Sbeta} and \eqref{H} it follows that we need to keep the coordinate dependence in the coefficient $D$ and in the thermodynamic potential $\Omega_s$. On the other hand, for what concerns the coefficients $C$ and $E$, it is fair to keep them equal to their bulk values $C(a_\infty)$ and $E(a_\infty)$.

\section{Results} \label{spatialprofiles}
In this section we present the numerical results for the soliton in the spin-polarized Fermi gas and study the effect of population imbalance on its properties. Moreover the range of applicability of the present effective field theory is determined and its dependence on temperature, imbalance and interaction is analyzed. 

\subsubsection{Shape of the soliton}

\begin{figure}[h!]\vspace{0.5cm}
\includegraphics[width=0.45\textwidth]{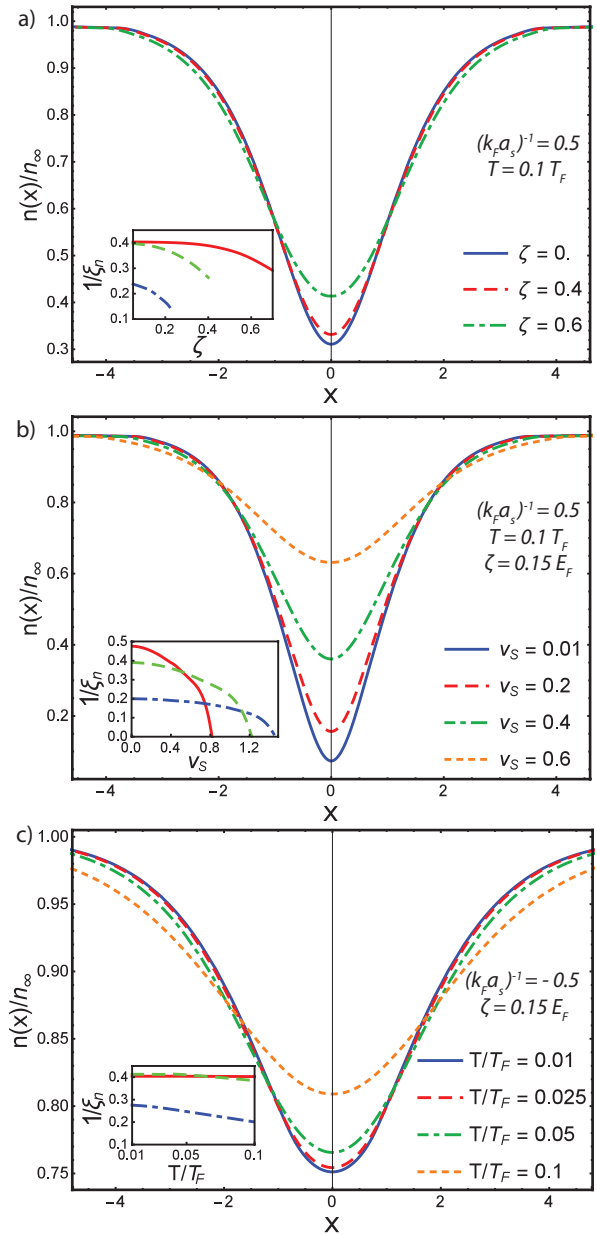}
\caption{(Color online) (a) Density profiles at $T=0.1T_F$, $v_S=0.4$ on the BEC side of the resonance $(k_Fa_s)^{-1}=0.5$ for different values of the imbalance parameter $\zeta$. (b) Density profiles at $T=0.1T_F$, $\zeta=0.15$ on the BEC side of the resonance $(k_Fa_s)^{-1}=0.5$ for different values of the soliton velocity. (c) Density profiles at $\zeta=0.15$, $v_S=0.4$ on the BCS side of the resonance $(k_Fa_s)^{-1}=-0.5$ for different values of the temperature. The insets show the behavior of the inverse soliton width $(\xi_n)^{-1}$ in the BCS (blue dotdashed line), unitarity (green dashed line) and BEC (red line) regimes as a function of $\zeta$, $v_S$ and $T$ respectively. The position $x$ and widths $\xi_n$ are given in units of $k_F^{-1}$, the imbalance parameter $\zeta$ is in units of $E_F$ and the velocities are in units of $v_F/2$.}\label{PRFn}\vspace{-0.5cm}
\end{figure}

\begin{figure}[h!]
\includegraphics[width=0.43401\textwidth]{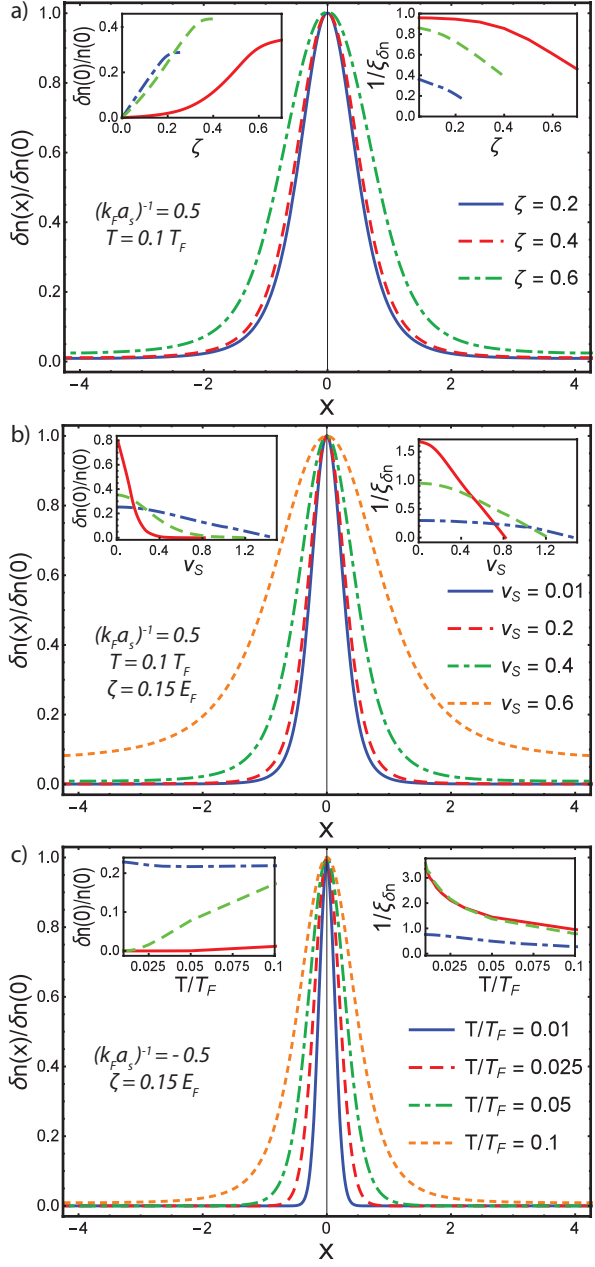}
\caption{(Color online) (a) Excess component density at $T=0.1T_F$, $v_S=0.4$ on the BEC side of the resonance $(k_Fa_s)^{-1}=0.5$ for different values of the imbalance. (b) Excess component density at $T=0.1T_F$, $\zeta=0.15$ on the BEC side of the resonance $(k_Fa_s)^{-1}=0.5$ for different values of the soliton velocity. (c) Excess component density at $\zeta=0.15$, $v_S=0.4$ on the BCS side of the resonance $(k_Fa_s)^{-1}=-0.5$ for different values of the temperature. The left (right) insets show the behavior of the ratio $\delta n (0)/n(0)$ (inverse width $(\xi_{\delta n})^{-1}$ of the excess component distribution) in the BCS (blue dotdashed line), unitarity (green dashed line) and BEC (red line) regimes as a function of the imbalance, soliton velocity and temperature respectively. The position $x$ and widths $\xi_{\delta n}$ are given in units of $k_F^{-1}$, the imbalance parameter $\zeta$ is in units of $E_F$ and the velocities are in units of $v_F/2$.}\label{PRFdn}
\end{figure}

\begin{figure}[] 
\includegraphics[scale=0.5]{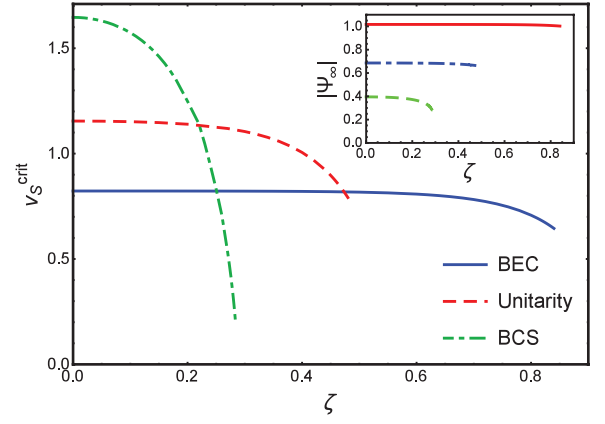}
\caption{(Color online) Critical velocity of the soliton as a function of the imbalance at $T=0.1T_F$ across the BEC-BCS crossover. The inset shows the corresponding behavior of $|\Psi_\infty|$. The imbalance parameter $\zeta$ is given in units of $E_F$, the velocities in units of $v_F/2$ and $|\Psi_\infty|$ in units of $E_F$.} \label{Vcrit}
\end{figure}

Figure \ref{PRFn}(a) shows the fermion density profile (normalized to the bulk density) of the soliton as a function of the distance from its center, for different values of the imbalance $\zeta$. The fermion density is calculated within the mean-field local density approximation (LDA), using the formula
\begin{equation}
n^{(\text{LDA})} = -\frac{\partial \Omega_s}{\partial \mu}\,.
\end{equation}
In \citep{THKTDPrA} the non-zero value of the density at the center of the soliton -- where, for low values of $v_S$, the order parameter approaches zero -- was explained by inferring that the soliton gets filled by unpaired fermions.
Consistently with this hypothesis we observe that the presence of imbalance enhances the value of the density at the soliton center.
More in general, for increasing values of $\zeta$, we can see that the soliton gets filled with a growing amount of particles, becoming less deep and slightly broader. This last effect is shown more clearly in the inset of Fig. \ref{PRFn}(a), which depicts the behavior of the inverse soliton width (taken at half the height of the density dip) $\xi_n^{-1}$.
While the density profile has been plotted only for the BEC side of the crossover, the inset shows that similar effects take place at unitarity and in the BCS regime. The values of the soliton width in the BCS regime are noticeably higher than those in the other cases.\\
Figures \ref{PRFn}(b)-\ref{PRFn}(c) also show the fermion density profile, but for different values of velocity $v_S$ and temperature $\slfrac{T}{T_F}$ respectively. The results for the temperature dependence are shown in the BCS regime where the effect of temperature is more evident.
The above-mentioned filling of the soliton for an increasing spin imbalance can be characterized more distinctly by calculating the density difference $\delta n (x)$ between the ``spin-up'' and ``spin-down'' populations along the soliton dip. In figure \ref{PRFdn}(a) we display the normalized $\delta n(x)/\delta n(0)$ profile for different values of the imbalance $\zeta$. The left inset clearly shows how the relative density difference at the center increases monotonously with the imbalance in all different regimes of the BEC-BCS crossover.
\\As the imbalance in the Fermi gas increases, so does the amount of unpaired particles that cannot participate in the superfluid state of condensated pairs. At higher temperatures, some of these normal state particles coexist with the condensate as a thermal gas, but any remaining excess of the majority component has to be spatially removed from the pair condensate. The soliton dip then turns out to be a suitable location to accommodate normal state particles, and consequently fills up with an increasing amount of unpaired particles as the imbalance gets higher.
As it is shown in the right inset, the width of the $\delta n(x)/\delta n(0)$ curves increases with $\zeta$ too, in agreement with the earlier observed broadening of the soliton density dip.\\
Similar to the density profile plots, we have also included figures of $\delta n(x)$ for different values of velocity $v_S$ and temperature $T/T_F$, presented in figure \ref{PRFdn}(b) and \ref{PRFdn}(c) respectively. In this regard it is interesting to notice that the decrease in temperature leads to an increasing degree of localization for the distribution of excess-spin component particles.\\From the insets of Fig.\ref{PRFn}(b) and Fig.\ref{PRFdn}(b) concerning the width $\xi_n$ of the soliton and $\xi_{\delta n}$ of the excess component distribution, it appears that a critical value of the soliton velocity can be determined above which the soliton cannot exist (the width of the profiles $n(x)$ and $\delta n(x)$ goes to infinity).
The effect of imbalance on this quantity has been analyzed and the results are shown in Fig.\ref{Vcrit} again all across the BEC-BCS crossover. From the comparison with the behavior of the mean field bulk value of the order parameter in the same regimes (inset) it becomes clear that the value of the imbalance for which the critical velocity $v_S^{(crit)}$ goes to zero is the same $\zeta^{(crit)}$ for which the minimum of the free energy corresponds to $|\Psi_\infty|=0$ i.e. when the normal state becomes energetically favorable over the superfluid one.

\subsubsection{Validity range of the treatment}

\begin{figure}[h]
\includegraphics[scale=0.5]{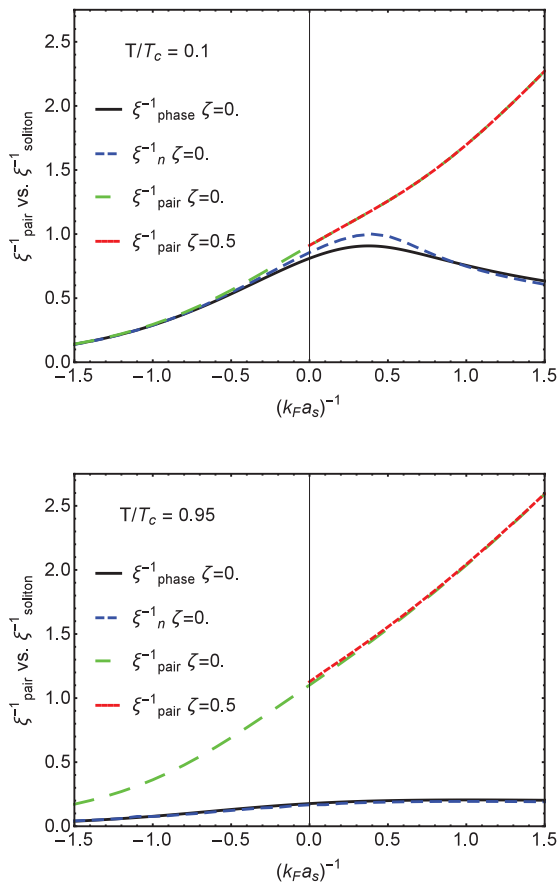}
\caption{(Color online) (a) Comparison between inverse pair coherence length and inverse soliton width (in units of $k_F$) across the BEC-BCS crossover for $T/T_c=0.1$. The full black line represents the inverse healing length without imbalance. The dashed blue line describes the inverse soliton width. The dashed green line represents the inverse pair coherence length. The dotted red line represents the inverse pair coherence length with imbalance $\zeta=0.5 E_F$. (b) Comparison between inverse pair coherence length and inverse soliton width across the BEC-BCS crossover for $T/T_c=0.95$: the dashing/color code is the same as for (a).}\label{figcoherence}
\end{figure}

\begin{figure}[h]
\includegraphics[scale=0.5]{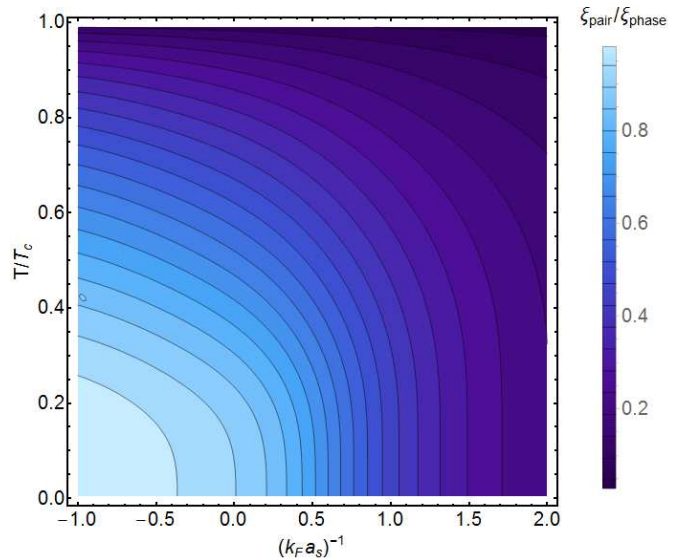}
\caption{(Color online) Contour plot depicting the behavior of the ratio $\xi_{pair}/\xi_{phase}$ between the pair coherence length and the variationally determined healing length as a function of the inverse scattering length $a_s$ and the temperature. The values of the scattering length span the entire BEC-BCS crossover, while the temperature range goes from $0$ to the critical temperature $T_c$.}\label{figpairhealing}
\end{figure}

\begin{figure}[h]
\includegraphics[scale=0.5]{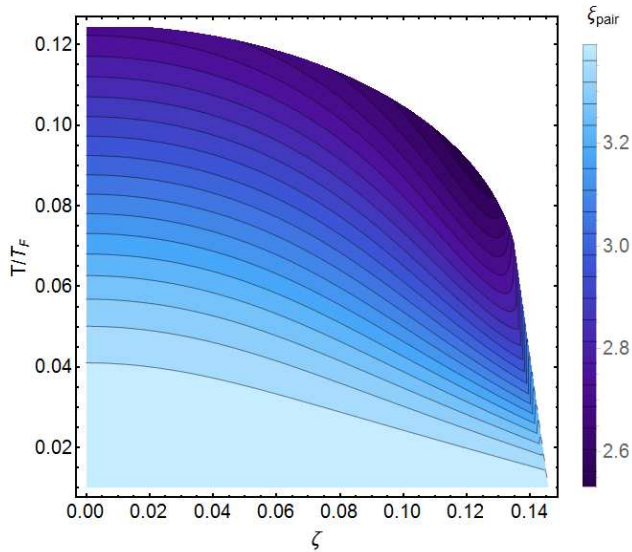}
\caption{(Color online) Dependence of the pair coherence length on the imbalance parameter $\zeta$ and temperature $T$ at $(k_Fa_s)^{-1}=-1$ on the BCS side of the resonance. The non-monotonic behavior of $\xi_{pair}(\zeta)$ is apparent. Lengths are in units of $k_F^{-1}$ and $\zeta$ is in units of $E_F$.}\label{figcontour}
\end{figure}

As mentioned above, the theoretical model employed in the present discussion is based on the assumption that the profile of the order parameter changes slowly in both time and space, corresponding to a long-wavelength approximation. In order to verify whether or not this assumption holds for the system under consideration, we compare the characteristic size of the soliton with the Cooper-pair correlation length $\xi_{pair}$ (often referred to as Pippard length \citep{THFetterWalecka}). The comparison ultimately enables us to determine the domain of validity of the EFT and identify it as the region of the $\left\{\beta,\zeta,(k_Fa_s)^{-1}\right\}$ -space for which the size of the soliton is much larger than $\xi_{pair}$.\\To describe the characteristic size of the soliton we will use (and compare) two different definitions: the width at half height of the soliton dip $\xi_n$ (already used in the previous section) and the healing length $\xi_{phase}$. In this context, $\xi_{phase}$ for the soliton at rest ($v_S=0$) has been calculated by modeling the spatial profile of the order parameter with a trial form of the amplitude modulation
\begin{equation}
a(x)=\tanh\left(\frac{x}{\sqrt{2}\xi_{var}}\right) \label{tanh}
\end{equation}
and then minimizing the free energy with respect to the variational parameter $\xi_{var}$. As discussed in \citep{THPalestiniStrinati} a rescaling coefficient has to be introduced in order to connect the variational parameter $\xi_{var}$ to the healing length $\xi_{phase}$. To this purpose, the convention of rescaling $\xi_{var}$ to the value of $\xi_{phase}$ at $T=0$ in the BCS limit was adopted, obtaining the relation $\xi_{phase}=1.175\,\xi_{var}$.\\The pair coherence length $\xi_{pair}$ has been calculated in \citep{THPistolesiStrinati,THPalestiniStrinati} in terms of the pair correlation function 
\begin{align}
g_{\uparrow\downarrow}(\bm{r})&=-\left(\frac{n}{2}\right)^2\\&+\left\langle\psi_{\uparrow}^\dag\! \left(\!\bm{R}+\frac{\bm{r}}{2}\right)\!\psi_{\downarrow}^\dag \!\left(\!\bm{R}-\frac{\bm{r}}{2}\right)\!\psi_{\downarrow} \!\left(\!\bm{R}-\frac{\bm{r}}{2}\right)\!\psi_{\uparrow}\! \left(\!\bm{R}+\frac{\bm{r}}{2}\right)\right\rangle
\nonumber\end{align}
by following the definition
\begin{align}
\xi_{pair}=\sqrt{\frac{\int\mathrm{d}\bm{r}\,\bm{r}^2g_{\uparrow\downarrow}(\bm{r})}{\int\mathrm{d}\bm{r}\,g_{\uparrow\downarrow}(\bm{r})}}.
\end{align}
At mean-field level, in the context of our effective field theory, this quantity can be obtained as
\begin{align}
\xi_{pair}=\sqrt{\frac{\int\mathrm{d}k\,k^2\left(4k\,\xi_{\bm{k}}f_2\left(\beta,E_{\bm{k}},\zeta\right)\right)^2}{\int\mathrm{d}k\,k^2\,\left(f_1\left(\beta,E_{\bm{k}},\zeta\right)\right)^2}}.
\end{align}
In Fig.\ref{figcoherence}a and Fig.\ref{figcoherence}b the behaviors of the inverse pair coherence length $\xi_{pair}^{-1}$, soliton width $\xi_n^{-1}$ and healing length $\xi_{phase}^{-1}$ in function of the interaction strength are compared for two values ($T/T_c=0.1$ and $T/T_c=0.95$ respectively) of the ratio between the temperature of the system and the critical temperature.
As expected a very good agreement is found between the variationally determined healing length and the soliton width $\xi_{n}$. Only in the vicinity of the unitarity regime we see a sizeable difference between these two quantities. This can be understood since, while in both the BCS \citep{THHeeger} and BEC \citep{THKonotopPitaevskii, THFrantzeskakis} limits theoretical predictions state that the soliton amplitude profile is described by a hyperbolic tangent as in \eqref{tanh}, at unitarity  and in the intermediate regimes this is generally not true.\\ For low values of the temperature (Fig.\ref{figcoherence}a), $\xi_{pair}^{-1}$ and $\xi_n^{-1}$ are very close in the unitarity and BCS regimes but deviate from each other when going towards the BEC side of the resonance, therefore limiting the validity domain of the EFT to this region. On the other hand, from the data sets relative to $T=0.95T_c$ (Fig.\ref{figcoherence}b) it is apparent that the soliton width remains noticeably higher than the Pippard length across the entire depicted window of the BEC-BCS crossover, guaranteeing the reliability of the predictions of the EFT in the whole domain. To have a better understanding of the validity range of the EFT, in Fig.\ref{figpairhealing} the ratio between the pair coherence length and the healing length is plotted as a function of both temperature (normalized to $T_c$) and interaction strength. The validity domain of the EFT can be intuitively identified with the dark blue/purple region in the contour plot.\\
For what concerns the dependence of $\xi_{pair}$ on $\zeta$, for the situations considered in Fig.\ref{figcoherence} we see that the presence of a population imbalance modifies just weakly the behavior of the pair coherence length. However, as shown in Fig.\ref{figcontour}, for certain values of $T$ a non-monotonic behavior of $\xi_{pair}(\zeta)$ is observed. This can be explained by examining the relation between the order parameter and the imbalance: in these configurations the modulus of the order parameter varies slowly in the entire range of values of $\zeta$ with the exception of the region in the immediate vicinity of the critical value $\zeta^{(crit)}$ where it suddenly drops to zero.
\subsubsection{Dynamical properties}

\begin{figure}[h]
\centering
\includegraphics[scale=0.55]{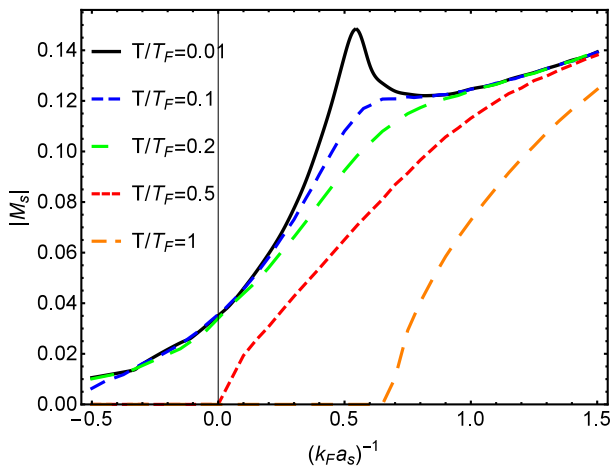}
\caption{(Color online) Absolute value of the effective mass of the soliton as a function of the interaction parameter $(k_Fa_s)^{-1}$ for different values of the temperature without imbalance ($\zeta=0$). The mass is given in units of $2m$.}\label{Meff0}
\end{figure}

\begin{figure}[h]\vspace{0.4cm}
\centering
\includegraphics[scale=0.55]{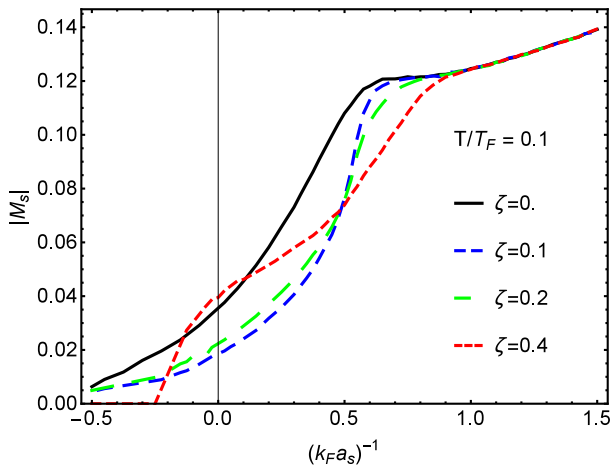}
\caption{(Color online) Absolute value of the effective mass of the soliton as a function of the interaction parameter $(k_Fa_s)^{-1}$ at temperature $T=0.1 T_F$ for different values of the imbalance (given in units of $E_F$). The mass is given in units of $2m$.}\label{MeffZ}
\end{figure}

\begin{figure}[h]
\includegraphics[scale=0.55]{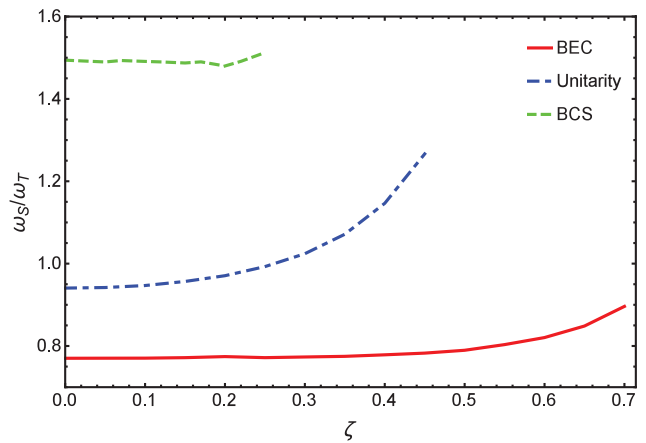} 
\caption{(Color online) Ratio of the soliton oscillation frequency $\omega_S$ to trapping frequency $\omega_T$, calculated as $\omega_S/\omega_T=\sqrt{ m N_s/M_S}$ as a function of the imbalance $\zeta$ (in units of $E_F$) at $T=0.1T_F$ and $v_S=0.6$ across the BEC-BCS crossover. Velocities are in units of $2 v_F$.}\label{TTtrap}
\end{figure}

For a balanced Fermi gas, it has been demonstrated that, within the effective field theory, a soliton obeys the energy-momentum relation of classic Hamiltonian dynamics
\begin{equation}
\frac{\partial \mathcal{E}_S}{\partial \mathcal{P}_S} = v_S \, ,
\end{equation}
i.e. the soliton behaves as if it were a classic particle.  Extending the formalism and calculations to a spin-polarized gas, we numerically confirm that this relation is still satisfied in the presence of imbalance. With this in mind, the effective mass $M_S$ of the soliton can be defined in relation to the soliton momentum $\mathcal{P}_S$ \eqref{Ps} and to the soliton energy $\mathcal{E}_S$ \eqref{Es} as in \citep{THFrantzeskakis}:
\begin{equation}
M_S \equiv \frac{\partial \mathcal{P}_S}{\partial v_S}\equiv \frac{1}{v_S}\frac{\partial \mathcal{E}_S}{\partial v_S} \, .\label{Mefff}
\end{equation}
The soliton's motion can then be treated as the motion of a classical particle with effective mass $M_S$ moving with velocity $v_S$.\\In Fig.\ref{Meff0} the behavior of the effective mass (which is a negative quantity) calculated at $v_S=0$ is examined across the BEC-BCS crossover for different values of the temperature. The effect of temperature becomes more evident in the interval of values of the interaction parameter ranging roughly between $0$ and $1$. For low temperatures ($T/T_F=0.01$) we notice the appearence of a peak centered around $(k_Fa_s)^{-1}=0.5$ while for higher temperatures a monotonic increase of the absolute value of $M_S$ is observed as we move from the BCS side towards the BEC side of the Feshbach resonance. The effect of imbalance on $M_S$ is shown in Fig.\ref{MeffZ} where the absolute value $|M_S|$ is plotted as a function of $(k_Fa_s)^{-1}$ for various values of $\zeta$. As for temperature, also in this case the change in the behavior due to $\zeta$ is more evident in the vicinity of the unitarity regime. While in the BEC and BCS limits we observe that $|M_S|$ is larger for small values of the imbalance and decreases with increasing $\zeta$ as could be intuitively expected in consideration of the discussion about the filling of the soliton by imbalance, in the intermediate region we see that the behavior is much less straightforward. In the immediate vicinity of $(k_Fa_s)^{-1}=0$ in fact, the higher value of $|M_S|$ is not reached for a balanced system but for a highly imbalanced one ($\zeta=0.4$).\\
A final remark has to be made for what concerns the influence of the imbalance on the period of oscillation of the solitonic wave in the trapped condensate. This period $T_s$ can be calculated according to the formula in \citep{THCetoliBrand}:
\begin{equation}
T_S = \sqrt{\frac{M_S}{m N_S}}T_T \,,\label{eq:Ts}
\end{equation}
where $T_T = \slfrac{2 \pi}{\omega_T}$ is the period associated with the frequency of the trap and $N_S$ is the (negative) amount of particles in the soliton dip (per unit area), defined as $\int_{-\infty}^{\infty}[n(x)-n(\infty)] \ dx$ . The quantity $m N_S$ represents the physical mass of the soliton, which in general is different from the effective mass. Since both the physical and effective mass of the dark soliton are negative, their ratio in (\ref{eq:Ts}) will be positive. It must be remarked that our present treatment, because of the assumption \eqref{solitondependence}, does not allow for oscillations of the soliton in the $(y,z)$ plane. Therefore the following calculations can be interpreted as a limiting case for a strongly confined (\textit{quasi-1D}) gas hence not allowing the snake instability to develop.\\
Figure \ref{TTtrap} shows the ratio of the soliton oscillation frequency to the trap frequency for the different regimes of the BEC-BCS crossover. In the BEC and unitarity regimes, the oscillation period increases monotonously for increasing values of $\zeta$. In the BCS regime, the changes in the oscillation frequency are very small compared to the other cases. 
As opposed to what is stated in \cite{THLiaoBrand,THLiaoBrandE,THScottDalfovo} -- and observed in \citep{EXPYefsah} in relation to solitonic vortices --, from Fig.\ref{TTtrap} we see an overall increase in the ratio $\omega_S/\omega_T$ as the system goes from the BEC towards the BCS regime. However, at low temperatures the dependence of $\omega_S/\omega_T$ on the interaction parameter is not monotonic: the ratio is found to first decrease from the deep BCS regime until it reaches a minimum for a value of $(k_Fa_s)^{-1}$ around $0.5$ and then to increase towards the theoretically predicted value of $1/\sqrt{2}$ as $(k_Fa_s)^{-1}$ reaches high positive values.
A sign of this discrepancy can be observed by considering the graphs relative to the soliton energy $\mathcal{E}_S$ in \citep{THKTDPrA} where the energy calculated in the framework of the present EFT is compared with the results obtained from the solution of the BdG equations. From this comparison it emerges that in both the BCS and unitarity regimes the slope of $\mathcal{E}_S(v_S)$ according to our theory is less steep than the BdG one. As it is clear from \eqref{Mefff}, this translates in an underestimation of the effective mass $M_S$ and, in turn, in an overestimation of the ratio $\omega_S/\omega_T$ in these regimes. To understand the observed difference between the predictions of the present EFT and the other results found in literature, three more elements have to be taken into account. First it has to be stressed that equation $\eqref{eq:Ts}$ is obtained by considering a soliton oscillating inside a trap, a condition that does not match our assumption \eqref{solitondependence} requiring the excitation to move at constant speed $v_S$ along the $\hat{x}$ direction. Also, due to the bosonic nature of the EFT, Andreev bound states cannot be straightforwardly included in the treatment: however their presence is predicted to give a sizeable contribution to the effective mass of the soliton \citep{THAntezzaDalfovo,THSpuntarelli,THScottDalfovo}. Finally in the present treatment we have fixed the value of the background density far away from the soliton $n_{\infty}=1/3\pi^2$: for a trapped system a correction should in principle be included in the calculation of $N_S$ to account for the variation of this quantity across the BEC-BCS crossover. 

\section{Conclusions} \label{conclusions}
In this paper we have analyzed the effect of temperature and population imbalance on the properties of dark solitons in ultracold Fermi gases across the BEC-BCS crossover. The study was performed in the context of an effective field theory illustrated in \citep{THKTLDEpjB} based on the only assumption of a slow-varying pairing field. The fact that no hypothesis was made requiring the pair field to be small, in principle extends the range of validity of this theory with respect to the widely employed Ginzburg-Landau and BdG approaches enabling us to consider also the effect of temperature on the system. To test the intrinsic reliability of the predictions of the EFT on dark solitons, a comparison has been made between the characteristic size of the excitation and the pair coherence length \citep{THPistolesiStrinati,THPalestiniStrinati}, resulting in a better understanding of how temperature and interaction affect the domain of applicability of the theory. In particular, while the calculations relative to the BEC regime are valid for all values of the temperature from $0$ to $T_c$, at unitarity and on the BCS side of the resonance this is true just for temperatures close to the critical one.\\
Based on analytic expressions for the amplitude and phase spatial profiles, the density and the density of the excess-spin component were obtained in LDA approximation using for the bulk value of the order parameter the mean field results. By systematically analyzing the density profiles we have observed how increasing the imbalance (and consequently decreasing the number of particles available for pairing) results in a filling of the soliton core that thus proves to be a convenient place where the unpaired particles can be stored. 
This translates into a decrease of the modulus of the effective mass of the excitation with increasing imbalance. However, in the crossover region in the vicinity of the unitarity regime we observe that the effect of the imbalance on $M_s$ is reversed.  A discrepancy was observed between our predictions concerning the change in the soliton-to-trap period across the BEC-BCS crossover and those reported in other papers all based on the solution of the time-dependent Bogoliubov-de Gennes equations \cite{THLiaoBrand,THLiaoBrandE,THScottDalfovo} at zero temperature. 
\\Keeping in mind the experimental setup employed in the investigation of solitons in ultracold gases a substantial part of our results was presented as a function of the soliton velocity $v_S$ so to have the possibility of a direct comparison with future experimental results.\\
\acknowledgements We gratefully acknowledge useful discussions with J.P.A. Devreese and N. Verhelst.
This research was supported by the Flemish
Research Foundation (FWO-Vl), project
nrs. G.0115.12N, G.0119.12N, G.0122.12N, G.0429.15N,
by the Scientific Research Network of the Research
Foundation-Flanders, WO.033.09N, and by the
Research Fund of the University of Antwerp.

\bibliography{Refs_experiment,Refs_theory}
\end{document}